CONQUERY: AN OPEN SOURCE APPLICATION TO ANALYZE HIGH CONTENT HEALTHCARE DATA

Fabian Kovacs[1], Max Thonagel[1], Marion Ludwig[1*], Alexander Albrecht[2], Manuel Hegner[2], Dirk Enders[1], Lennart Hickstein[1], Maximilian von Knobloch[1], Anne Rothhardt[1], Jochen Walker[1]

[1]InGef – Institute for applied health research Berlin GmbH, Germany

[2]bakdata GmbH, Berlin, Germany

*Corresponding author:

Marion Ludwig

Spittelmarkt 12, 10117 Berlin

Marion.ludwig@ingef.de

+49 30 586945 473

*ABSTRACT*

**Introduction:** Big data in healthcare must be exploited to achieve a substantial increase in efficiency and competitiveness. Especially the analysis of patient-related data possesses huge potential to improve decision-making processes. However, most analytical approaches used today are highly time- and resource-consuming.

**Objectives:** The presented software solution Conquery is an open-source software tool providing advanced, but intuitive data analysis without the need for specialized statistical training. Conquery aims to simplify big data analysis for novice database users in the medical sector.

**Methods:** Conquery is a document-oriented distributed timeseries database and analysis platform. Its main application is the analysis of per-person medical records by non-technical medical professionals. Complex analyses are realized in the Conquery frontend by dragging tree nodes into the query editor. Queries are evaluated by a bespoke distributed query-engine for medical records in a column-oriented fashion. We present a custom compression scheme to facilitate low response times that uses online calculated as well as precomputed metadata and data statistics.

**Results:** Conquery allows for easy navigation through the hierarchy and enables complex study cohort construction whilst reducing the demand on time and resources. The UI of Conquery and a query output is exemplified by the construction of a relevant clinical cohort.

**Conclusions:** Conquery is an efficient and intuitive open-source software for performant and secure data analysis and aims at supporting decision-making processes in the healthcare sector.

*Introduction*

Improved methodology in medical imaging, accumulation from research-related data such as individual multiomics analyses, health apps and wearable devices (Internet of Medical Things) contribute to an estimated growth of health data by an annual 48%. [1] For the challenge of combining or analysing these heterogenous data sources in interaction with the existing structures of clinical and claims data, generalisable tools are needed.

Interestingly, unlike other industries the German healthcare sector often neglects the connection between cost efficiency and data analytics. A cooperative study of the German health market by McKinsey and the BMV e.V. put a number on digitalization and figured a potential saving of 34 billion Euro in 2018, of which 70% could have arisen at the site of care providers. [2] In practice, varying standards in data generation, editing and storage, as well as requirements regarding data protection and security are technical obstacles preventing broad usage of clinical data. In a first step this should be tackled by a flexible clinical data warehouse (DWH) infrastructure. Key challenges of a clinical DWH are integration of heterogenous data sources and their standardization, a proper extract, transform and load process as well as security and privacy provisions. Following the establishment of a central DWH, real-world data should be used to increase efficiency and to improve healthcare for all parties involved.

In addition to technical challenges, the immediate work on relevant research questions using the large data treasure often fails because of a fixed data sovereignty or due to the inability to analyze large data volumes. This is true for most data sources, such as clinical data, claims data or laboratory data, so that the generated data remains unused in many cases. Important research questions usually arise among medical personnel, who, however, is often limited in time and advanced statistical knowledge and database experience. IT personal who could support the process of big data analysis are either not available or communication is too time-consuming to deal with topics promptly. Further, stringent requirements regarding data privacy protection or limited financial resources render data mining efforts ineffective. Regarding the German healthcare market, healthcare providers and insurance funds could opt between proprietary platforms or in-house IT developments to analyze their generated and stored data. However, open source platforms offer many advantages over commercially available analysis tools. Most importantly, they act as a starting point for sound data analysis and at the same time provide the flexibility for the development of own, specialized plugins. [3], [4] Technical service and support usually come from a dedicated community that continuously incorporates customized and adaptable tools. Further, open source platforms are hoped to encourage precompetitive data sharing based on open (and joint) standards.

*OBJECTIVES*

We developed Conquery, an open source analytical application [5] that could complement any DWH to address the need for an easy-to-use analytical software in the medical sector. Using Conquery, complex queries are realized with intuitive drag-and-drop functions and without the need for specialized statistical training. The design objectives of Conquery are to offer a secure platform for (patient) data analysis and to provide a software with an easily expandable framework. Currently, its main application is the analysis of per-person medical records by non-technical medical professionals aiming to support decision-making and transparency. In contrast to other open source solutions, such as i2b2, [3], [4], Conquery allows for performant query processing and cohort selection and analysis against very large datasets. This paper provides a first technical description of the analysis software Conquery.

For the sake of comprehensibility, we will differentiate between operators and users. Operators are responsible for maintaining a Conquery instance. They define schemas, concepts and import the data, hence technical knowledge (JSON, SQL etc.) and domain expertise is required. Users are the non-technical medical experts, who build queries through the browser-based user interface to perform an analysis. The data organization of Conquery is illustrated schematically in figure 1.

*Architecture overview*

Conquery is implemented in Java, using the manager-worker paradigm for distributed parallel computation. The manager is responsible for distributing data among the workers and relaying incoming user-queries to the workers. The workers hold data partitioned by unique identifiers and will evaluate incoming queries, allowing Conquery to be run on a single machine or scaled to a distributed cluster. As will be outlined below, queries are evaluated on a per person basis, enabling trivial parallelization of query evaluation. Referring to per person basis is technically interchangeable against any chosen identifier and exemplifies the architecture of Conquery.

*Tables & Imports*

Tables define a data-scheme composed of columns and their data-type descriptions (text, money, number, date, etc.), serving as the building blocks of so-called Concepts (compare section Concepts). Data can be loaded into a table after it has undergone a mandatory preprocessing step where it is analyzed, structured, and then compressed using common techniques on a per column basis: Integers are bit-packed [6], text undergoes dictionary-encoding where appropriate [7], and other types are where possible reduced into integer representations that also undergo bit-packing.

The data is kept in memory, in column-oriented fashion, in this compressed representation and is only de-compressed when necessary for evaluation (e.g.: equality checks on text), adding overhead to query performance but reducing memory utilization significantly. Furthermore, we leverage the fact that the Java Virtual Machine uses just-in-time compilation to generate optimized code according to its usage (e.g.: inlining method calls, pruning unused code, removing redundant checks and applying auto-vectorization). [8]

This compression yields an average size reduction of 82% compared to raw data. Applying gzip on the same files achieves a similar compression ratio of 81%. However, Conquery's compression scheme allows for in-memory seeking single event precision which gzip does not support natively. [9]

*Concepts*

Concepts are the central component of Conquery. Here, table filters and aggregations are defined for users, enriched with descriptions. In addition, Concepts allow automatic classification of data according to an operator defined hierarchical schema, such as ICD-10-GM coding, de-structuring it onto a tree-like structure.

To exemplify this, the ICD-10-GM code "G20.11" for the diagnosis "Parkinson's disease with fluctuations and medium to severe impairment" is used and illustrated in figure 3. It is first classified as "G00-G99 - Diseases of the nervous system", subordinated "G20-G26 - Extrapyramidal and movement disorders", followed by "G20 Parkinson's Disease", and then "G20.11 - Parkinson's disease with fluctuations and medium to severe impairment". This hierarchy is defined in a tree-structured JSON-file (see supplementary data code #1). A query for all diagnoses of "G20-G26" will therefore not only contain "G20" diagnoses, but also every subordinated diagnosis.

Classification can also undergo more involved conditions, such as basic Boolean logic and filtering along additional columns. As this is a potentially complex query, we pre-compute assignments at load time, optimizing them further using Trie-data-structures as index structure and caching layers. [10] The data in Conquery is organized as tables, imports and concepts is illustrated schematically in figure 1.

*Queries, Filters and Aggregations*

The basic semantics of a Conquery query are "return all entities, satisfying X", where X is a set of conditions on the entities' records. Additionally, every query returns an aggregation of the times over which *X* holds true for the entity, and if specified, aggregation results over the entity. We also allow more fine-grained analysis grouping the records by a secondary key, for example to analyze links between patients and care providers or hospitalization cases. The lifetime of a query is illustrated and described in detail in figure 2. Since computation is independent per entity, we can trivially parallelize the query-engine, achieving high throughput performance.

The operators define filters and aggregations in the concepts to be used by users (examples provided in supplementary data code #2). They either limit the incoming records to a subset fulfilling certain criteria (e.g.: events happening in a certain time frame, made by a certain specialist, or a value being in a certain range), or aggregate over the filtered records and limit the outgoing entities based on user defined criteria over those aggregations (e.g.: having less than X diagnoses, or at most Y number of days in hospital care). These aggregations can also be provided as output for the user.

An incoming query consists of the above outlined filters and operators (see supplementary data code # 3). These can be optionally combined by logical operators (i.e.: and, or, not). The query is forwarded to all workers, where it is first translated into an optimized query plan for execution, and then evaluated for each entity.

Since actual query computation is expensive, we maintain indices and statistics of every partition and entities in relation to the concepts. We use those to only evaluate the table-partitions that contain relevant information for the requested concepts. Contained entities are batched and sent to the manager node with auxiliary data (the aggregations the user requested). Finally, a result file (CSV, Excel, Arrow [11]) is issued to the user (figure 2, step 8).

*RESULTS*

Conquery was developed to provide an intuitive platform for the prompt analysis of upcoming medical hypotheses. Especially important steps as immediate cohort formation for clinical studies or for the evaluation of a treatment method is currently limited by the dependence on statistically trained personal or insufficient statistical programs.

The user interface of Conquery is designed to displays concepts according to the outlined tree structure, in a directory like format, allowing novice users easy navigation through the hierarchy (figure 3). Queries can be composed by dragging tree nodes onto the query editor, where the user can choose to combine (OR) or intersect (AND) a node-criteria with further nodes by choosing the corresponding AND/OR-drop-zones (compare Fig. 3). Clicking on a query element opens an advanced settings dialogue box to define further parameters (figure 4). This allows users to select specific data sources or extending the output to contain further aggregation results through the options found under "additional values". However, the extent and content of the dialogue field depends on the underlying node concept. The query can be executed by clicking the "start query" button and a "CSV" button will appear when the backend has finished query execution. This offers the option to download the resulting CSV file for further data processing (figure 5). The query is automatically saved as a query history entry and can be used in later queries. For advanced users Conquery offers an API (Python/HTTP) to programmatically formulate queries. This can also be leveraged for the integration of standardized analyses using other statistic backends.

A real clinical case could therefore be queried according to figure 3. For a required cohort of patients with Parkinson disease, only patients with a main discharge diagnosis (inpatient) OR two confirmed diagnoses within two quarters (inpatient and outpatient) OR with a Parkinson disease diagnosis made by at least two physicians OR a Parkinson disease diagnosis and at least one prescription of a Parkinson disease relevant medication were included. Moreover, those patients were eligible if they were continuously insured in the investigated time period. For each concept time periods can be defined, and a negation of the respective concept is possible. Figure 4 illustrates the advanced setting dialogues box available and customizable for each concept. Clicking on the query element obtained via the ICD concepts, ICD diagnosis made in the outpatient (physician) and inpatient (hospital diagnoses) sector can be distinguished. For the Parkinson cohort, patients were eligible if they had at least two diagnoses (cases) within at least two quarters in the inpatient sector. Further, the "additional values" can be chosen to extract more detailed information on each respective case of the cohort. Once the inclusion and exclusion criteria of the cohort are set, the query can be executed resulting in a readable and editable CSV format, which can be used for further analysis or graphical representations. All selected "additional values" can be found as separate columns in the CSV (figure 5), i.e., a list of ICD-codes is output for each patient in the cohort. Accordingly, this cohort can be used as a basis for an upcoming clinical study or might support evaluations of resource consumption of specific patient groups.

*DISCUSSION*

*Principal Results*
Conquery serves as the basis for an analysis tool developed by InGef that is used by various German statutory health insurance providers and is currently being evaluated for use by various hospitals and research institutions. Once implemented in a local DWH-setup, it could be used for real world evidence studies and generation of hypotheses, to improve quality of care and study start-up or to enhance patient recruitment and the execution of clinical trials. The analytical possibilities grow with the extension of the integrated data in the specific DWH. In a next step, Conquery aims to support research institutions in understanding and efficiently using their obtained data and thus, to increase data value and the feasibility of prospective projects. Future work will encompass cloud integration and scalability features, preview and visualization features for the frontend and further performance improvements of the query engine. Through the active development and intensive involvement of users, major and minor features not listed are expected to be added. Providing the recent interface standard FHIR HL7 is a prospective challenge and will further improve software usability. Moreover, Conquery offers the import and integration of heterogenous data types, such as data from claims data analysis, health apps, medical registries, or electronic medical records.

*Limitations*
The design of Conquery is single-pass and entity-independent query evaluation, this results in two major limitations of Conquery's query engine. The query language does not allow for relative queries (e.g.: a person being in the 50% percentile of a ranking) or SQL-like joins. Both justified by the objective to avoid inter-worker communication (reducing programmatic complexity and query latency). To achieve similar functionality, we instead offer APIs (eg.: https://github.com/ingef/cqapi) for implementing new query logic to coordinate multiple queries.

*Conclusions*

New findings from basic research, health service research or clinical studies should be brought to the bedside at a faster pace than recently observable. At the same time, clinical routine must become more efficient to meet current and upcoming challenges in the healthcare sector, as fewer resources combined with more patients and more complex conditions increase the pressure in the medical sector. We developed a novel user-friendly analytical tool, which is provided open source at www.conquery.org. Conquery aims to add value to data by simplifying data analysis. Intuitive drag-and-drop functions enable even novice users to analyze large data sets with state-of-the-art methods.

AVAILABILITY AND REQUIREMENTS

Project name: Conquery
Project home page: www.conquery.org
Operating system(s): Linux/Windows
Programming language: Java
Other requirements: Java 11 or later
License: MIT

*DECLARATIONS*

*Acknowledgments*

We thank Peter Ihle from the working group "Erhebung und Nutzung von Sekundärdaten" (AGENS group of DGSMP and DGepi) for providing a test dataset for a demo version of Conquery.

Current affiliation of Manuel Hegner is tarent GmbH.

*Availability of data and materials*

Not applicable.

*Conflicts of interests*

None declared.

*Abbreviations:*

DWH – data warehouse

InGef – Institute for applied health research Berlin GmbH

*Funding*

Partly funded by the BMWi project HLaN-Health Reality Lab Network (01MD18004D)

*Author Contributions*

JW: original idea, led study, manuscript drafting and editing; ML/DE: manuscript drafting and editing; MH: software development; AA: manuscript drafting and editing; FK: software development, manuscript drafting and editing; MvK/LH/MT/AR/HP: software development and support for manuscript discussions. All authors read and approved the manuscript.

**Figure 1 Data organization of Conquery**

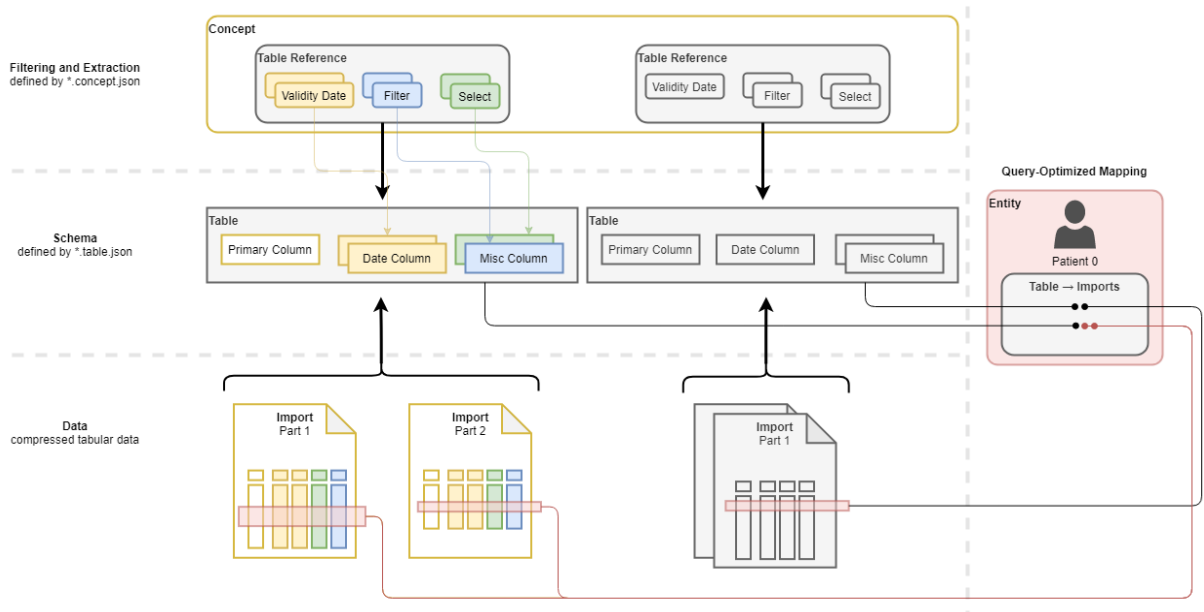

**Figure 2 Lifetime of a query in Conquery**

1. Submission of an API-layer description of the query, by the user, to the manager node 2. Distribution of the Query description to all worker nodes 3. Translation of the API-description to a concrete query plan (Optimization of the Query and preparation of necessary data-structures) 4. Per-Entity submission of a job into a queue for parallel processing 5. Evaluation of the query per entity, producing at least one result line for entities which satisfy the query, but can produce more for specific types of queries (outlined below) 6. Batched transmission of query results to the manager node 7. If all worker nodes have sent results, or a single node has sent a failure-result, the query is marked as finished and the user can download the result in CSV format if available 8. Rendering of the query to CSV from internal representation.

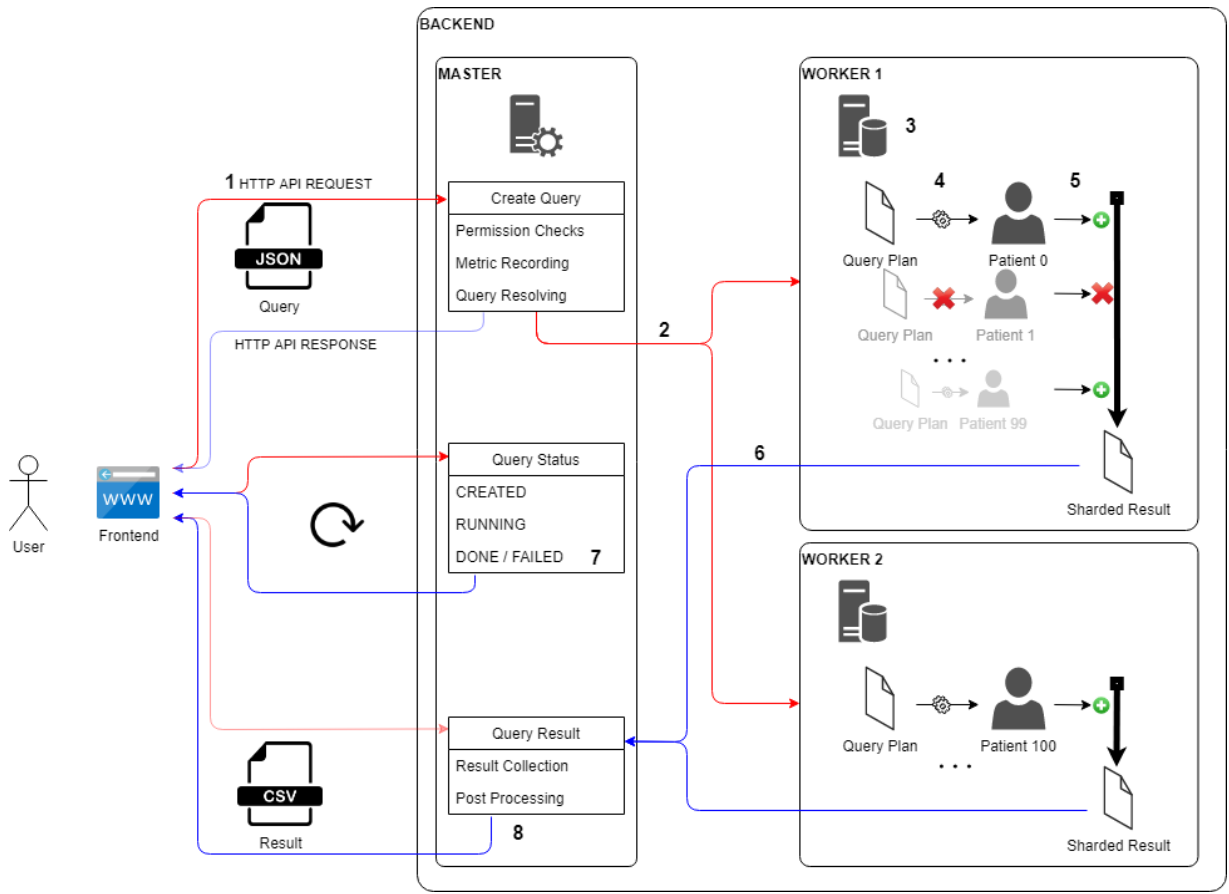

**Figure 3 Conquery user interface to construct queries.**

Specific concepts are dragged from the „CONCEPTS" panel into the „EDITOR" panel (right). The user can choose between a „GROUP EDITOR" (setup of the concept-based study cohort) or a „TIMEBASED EDITOR", which takes into account the chronology of events. The resulting cohort can be accessed via an output CSV file or reused for further queries. Previous queries are available in the „QUERIES" panel (middle). Basic information on the concepts can be found in the left „INFO" panel.

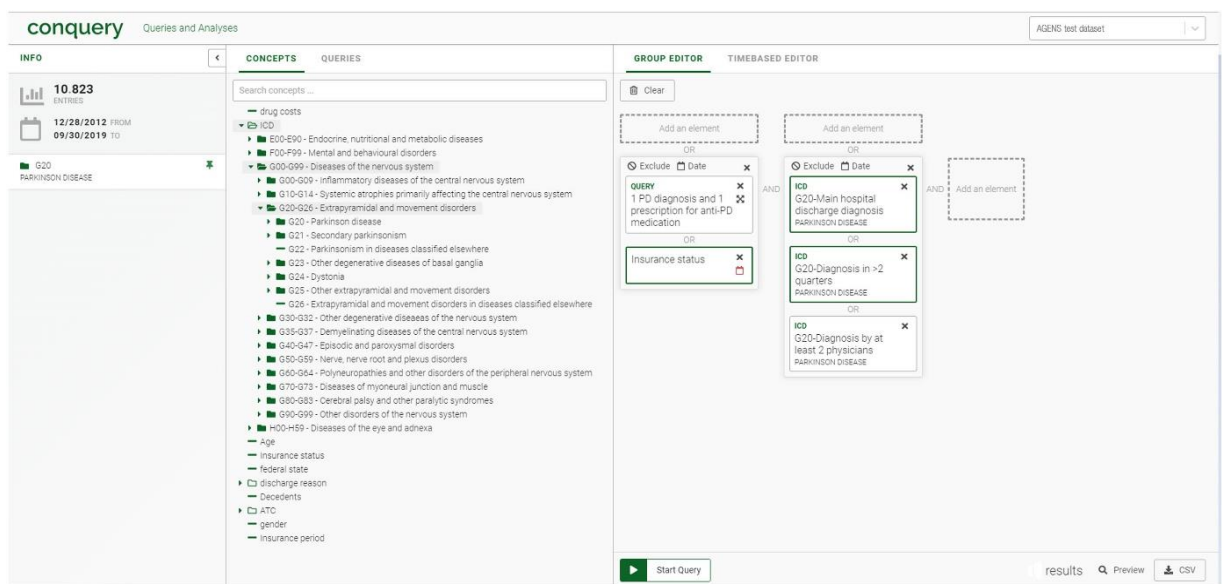

**Figure 4 Exemplary advanced settings dialogue box.**

Clicking on a concept in the editor leads to an extended settings box whose interface depends on the respective concept. Shown here is the corresponding box for an ICD concept (G20 – Parkinson disease). In this box you can distinguish between settings for inpatient (hospital diagnoses) and outpatient visits (physicians). Furthermore, special additional values ("Zusatzwerte"), such as List of ICD-diagnoses and Number of Cases can be selected, as well as the study population can be further defined by applying certain criteria. Shown here is the criterion ICD code at least in 2 quarters and at least 2 hospital cases per included patient. This enables the user to clearly define the study population. *Translation of German terms: Eigenschaften, characteristics; Zusatzwerte, additional values; Relevanter Zeitraum, relevant time period; Datumsspalte, date column; Bereich, section; Auswählen, select.*

**Figure 5 Exemplary csv output**

For the sake of clarity, a limited number of included patients (PID: 1-11) are shown. The csv document displays the requested variables for the individual cases. In the example, the date/data of the requested ICD diagnosis and all selected additional values are displayed in individual columns.

| PID | date | Outpatient ICD-Code | Number of physician visits | Number of quarters | Number of physicians visited | List of inpatient ICD-Codes | Number of hospitals visited | Length of hospital stays |
|---|---|---|---|---|---|---|---|---|
| 1 | {2015-07-16/2015-07-16, 2015-12-04/2015-12-04} | - | - | - | - | [G2090, G2000] | 2 | 10 |
| 2 | {2015-01-01/2015-12-31} | [G2090] | 11 | 4 | 3 | - | - | - |
| 3 | {2015-07-01/2015-12-31} | [G2090] | 2 | 2 | 1 | - | - | - |
| 4 | {2015-01-01/2015-12-31} | [G2090, G2010] | 12 | 4 | 3 | [G2090] | 1 | 4 |
| 5 | {2015-02-05/2015-02-05, 2015-02-10/2015-02-10} | - | - | - | - | [G2010] | 2 | 8 |
| 6 | {2015-04-01/2015-12-31} | [G2090, G2010] | 4 | 3 | 2 | - | - | - |
| 7 | {2015-01-01/2015-03-31, 2015-09-18/2015-09-18} | [G2090, G2000] | 2 | 2 | 2 | [G2090] | 1 | 10 |
| 8 | {2015-01-01/2015-12-31} | [G2090, G201] | 9 | 4 | 5 | [G2010, G2020] | 2 | 33 |
| 9 | {2015-01-01/2015-12-31} | [G2090] | 5 | 4 | 2 | - | - | - |
| 10 | {2015-01-01/2015-12-31} | [G2090] | 9 | 4 | 3 | [G2090] | 1 | 4 |
| 11 | {2015-01-01/2015-12-31} | [G2090, G2000, G2010] 11 | 4 | 4 | [G2010] | 1 | - |
| ... | | | | | | | | |

## Supplementary Data

Code #1 Example of a concept tree

```
{
    ...
    "name": "g20-g26",
    "description": " Extrapyramidal and movement disorders",
    "condition": {
        "type": "PREFIX_RANGE",
        "min": "G20",
        "max": "G26"
    },
    "children": [
        {
            "name": "g20",
            "description": " Parkinson's disease",
            "condition": {
                "type": "PREFIX",
                "prefix": "G20"
            },
            "children": [
                {
                    "name": "g20_1",
                    "description": " Parkinson's disease with moderate to severe impairment ",
                    "condition": {
                        "type": "PREFIX",
                        "prefix": "G201"
                    },
                    "children": [
                        {
                            "name": "g20_11",
                            "description": " Parkinson's disease with moderate to severe impairment with fluctuations",
                            "condition": {
                                "type": "PREFIX",
                                "prefix": "G2011"
                            }
                        }
                    ]
                }
            ]
        }
    ]
}
```

Code #2 Example of filters and selects

```json
{
    "label": "Hospital Diagnoses",
    "validityDates": [
        {
            "label": "Case begin",
            "column": "hospital_diagnosis.case_begin"
        },
        {
            "label": "Case end",
            "column": "hospital_diagnosis.case_end"
        }
    ],
    "column": "hospital_diagnosis.icd_code",
    "filters": [
        {
            "type": "SELECT",
            "label": "Diagnose kind",
            "column": "hospital_diagnosis.kind",
            "labels": {
                "primary": "Primary",
                "secondary": "Secondary",
                "initial": "Initial"
            }
        },
        {
            "type": "COUNT",
            "distinct": true,
            "label": "Case number",
            "column": "hospital_diagnosis.case_id"
        }
    ],
    "selects": [
        {
            "label": "ICD-Codes",
            "type": "DISTINCT",
            "column": "hospital_diagnosis.icd_code"
        },
        {
            "label": "Number of Cases",
            "type": "COUNT",
            "distinct": true,
            "column": "hospital_diagnosis.case_id"
        }
    ]
}
```

Code #3 Example of a simple query using filters and selects

```json
{
    "type": "CONCEPT_QUERY",
    "root": {
        "type": "CONCEPT",
        "ids": [
            "dataset.icd.g00-g99.g20-g26.g20"
        ],
        "tables": [
            {
                "id": "dataset.icd.hospital_diagnoses",
                "filters": [
                    {
                        "type": "SELECT",
                        "filter": "dataset.icd.hospital_diagnoses.diagnosis_kind",
                        "value": "primary"
                    }
                ],
                "selects": [
                    "dataset.icd.hospital_diagnoses.number_of_cases"
                ]
            }
        ]
    }
}
```